\documentclass[12pt]{article}
\usepackage{amsfonts}
\usepackage{latexsym}
\usepackage{amsmath}
\usepackage{amssymb}
\usepackage{amssymb}

\newcommand{\newsection}{    
\setcounter{equation}{0}\section}
\def\appendix#1{\addtocounter{section}{1}\setcounter{equation}{0}
\renewcommand{\thesection}{\Alph{section}}
\section*{Appendix \thesection\protect\indent \parbox[t]{11.15cm}{#1}}
\addcontentsline{toc}{section}{Appendix \thesection\ \ \ #1}}

\newcommand{\be}{\begin{eqnarray}}
\newcommand{\ee}{\end{eqnarray}}
\newcommand{\bea}{\begin{eqnarray}}
\newcommand{\eea}{\end{eqnarray}}
\newcommand{\ba}{\begin{array}}
\newcommand{\ea}{\end{array}}
\newcommand{\nn}{\nonumber \\}

\newenvironment{frcseries}{\fontfamily{frc}\selectfont}{}

\def \la {\label}

\def\a{\alpha}
\def\b{\beta}

\def\g{\gamma}

\def\e{\epsilon}

\def\bbe{{\bf{e}}}
\font\mybb=msbm10 at 11pt

\def\bb#1{\hbox{\mybb#1}}

\def\bR {\bb{R}}

\def\hn {{\tilde{\nabla}}}

\begin{document}
\begin{titlepage}
\begin{center}
\vspace*{-1.0cm}

\vspace{2.0cm} {\Large \bf  Static M-horizons} \\[.2cm]

\vspace{1.5cm}
 {\large  J. Gutowski and  G. Papadopoulos}

\vspace{0.5cm}
Department of Mathematics\\
King's College London\\
Strand\\
London WC2R 2LS, UK\\

\end{center}

\vskip 1.5 cm
\begin{abstract}
We determine the geometry of all static black hole horizons
of M-theory preserving at least one supersymmetry. We demonstrate that all such horizons
 are either warped products $\bR^{1,1} \times_w {\cal S}$ or $AdS_2\times_w {\cal S}$,
 where ${\cal S}$ admits an appropriate $Spin(7)$ or $SU(4)$ structure respectively;
 and we derive the conditions imposed by supersymmetry on these structures.
 We show that for electric static horizons with $Spin(7)$ structure, the near horizon geometry
 is a product $\bR^{1,1} \times {\cal S}$, where $\cal{S}$ is a compact $Spin(7)$ holonomy manifold.
 For electric static solutions with $SU(4)$ structure, we show that
 the horizon section ${\cal S}$ is a circle fibration
 over an 8-dimensional K\"ahler manifold which satisfies an
 additional condition involving the Ricci scalar and the length of the Ricci tensor. Solutions include
$AdS_2\times S^3\times CY_6$  as well as many others constructed from taking the 8-dimensional
K\"ahler manifold to be  a product of K\"ahler-Einstein and Calabi-Yau spaces.
\end{abstract}

\end{titlepage}


\section{Introduction}

The classical uniqueness theorems for 4-dimensional black holes \cite{israel}-\cite{robinson} do not extend to higher dimensions. In particular, in 5 dimensions, apart from black holes with spherical horizon topology \cite{bmpv} there are also black rings with horizon topology $S^1\times S^2$ \cite{ring1}. In more than 5 dimensions, the results of \cite{gibbons1, rogatko, reall3, obers1} indicate
that there are many black holes with exotic horizon topologies.

The question naturally arises as to whether there are black holes with exotic horizon topologies in
10- and 11-dimensional supergravities, which arise as effective theories of strings and M-theory.  For this a near horizon analysis has been carried out in the heterotic \cite{hh} and IIB \cite{iibh}
supergravities. This analysis  has led to the discovery of many new black hole near horizon geometries, and so has provided some supporting evidence for the existence of exotic black holes in these theories.

In this paper, we shall investigate the static near horizon black hole geometries of 11-dimensional supergravity which preserve at least one supersymmetry. It is expected that there are many black hole solutions
in M-theory. The IIA Newton constant increases quadratically with the string coupling.
So as the IIA string coupling becomes large, the strength of the gravitational force increases
and IIA matter collapses to black holes. But the strong coupling limit of IIA string theory
is conjectured to be M-theory \cite{townsend, witten} which has as an effective theory 11-dimensional supergravity. So these black holes should be solutions of 11-dimensional supergravity \cite{julia}.

As in the case of heterotic and IIB black hole horizons the aim is to find
all black hole horizons of 11-dimensional supergravity which preserve one supersymmetry.
However, unlike the heterotic case, there is no complete classification of supersymmetric backgrounds
in 11-dimensions. The Killing spinor equations (KSEs) of 11-dimensional supergravity for backgrounds preserving
one supersymmetry have been solved in \cite{pakis}, and in \cite{spingem} using spinorial geometry. A systematic method
for solving the KSEs of 11-dimensional supergravity for backgrounds preserving any fraction of supersymmetry has been presented in  \cite{system11}. Moreover it has been shown that all backgrounds which preserve more than
29 supersymmetries are maximally supersymmetric \cite{ggp}, and the maximally supersymmetric backgrounds
have been classified in \cite{maxsusy}. There are also conjectures on the number of supersymmetries preserved by supersymmetric M-theory backgrounds  \cite{duff} and the geometry of solutions with more than 24 supersymmetries \cite{jose2}.

The focus of the work will be on the static near horizons of black holes which preserve at least one supersymmetry. The addition
of rotation makes the analysis more involved and it will be reported elsewhere. Facilitated by the spinorial geometry technique for
solving KSEs of \cite{spingem, system11},
we show that the solution of the   KSEs of 11-dimensional supergravity  imply that the
near horizon geometries preserving at least one supersymmetry are either warped products $AdS_2\times_w {\cal S}$,
or {\it locally} are warped products $\bR^{1,1} \times_w {\cal{S}}$, where ${\cal S}$ is the near horizon section which admits either a  $SU(4)$  or a $Spin(7)$
structure, respectively.
In both cases, we present all the geometric conditions on the $SU(4)$ and $Spin(7)$ structures implied by the KSEs. In the former case,
 if ${\cal S}$ admits an isometry, then it is a fibration  over an almost Hermitian symplectic 8-dimensional
base manifold $B$. The  skew-symmetric part of the Nijenhuis tensor of $B$ vanishes but the almost complex structure is not always integrable.
The conditions on the geometry we have found and the $AdS_2$ backgrounds we have considered are {\it more general than} those that have appeared so far in the literature \cite{kim, kim3, kim2} in the context
of $AdS_2$ solutions in
11-dimensional supergravity.

The field equations impose additional conditions. We have solved these for the electric static horizons.
The electric near horizon geometries with a $Spin(7)$
structure are products $\bR^{1,1}\times {\cal S}$, where ${\cal S}$ is a holonomy $Spin(7)$ manifold, and the 4-form flux vanishes.
In  the $SU(4)$ structure case, ${\cal S}$ admits an
isometry and  is  a fibration over  an 8-dimensional K\"ahler manifold $B$. In addition, the Ricci scalar and the length of the Ricci tensor
of the K\"ahler manifold satisfy a condition (\ref{rrr}). This condition has been
previously found in   the context of
AdS${}_2$/CFT${}_1$ correspondence \cite{kim}. In the special case where the solution is a direct
product $AdS_2\times {\cal S}$, ie the warp factor is constant, the Ricci scalar is constant and so
$B$ is a K\"ahler-Yamabe manifold. Furthermore the length of the Ricci tensor is also pointwise
constant.  It turns out that $B$ is not a K\"ahler-Einstein space, and  ${\cal S}$ is not Sasakian.
Solutions include $AdS_2\times S^3\times CY_6$  for any 6-dimensional Calabi-Yau manifold $CY_6$ and
others which can be constructed by taking $B$ to be a product of K\"ahler-Einstein and Calabi-Yau manifolds.
Such solutions have also been found  in \cite{kim, kim2} searching for backgrounds in the context
of AdS${}_2$ solutions in 11-dimensional supergravity.

This paper is organized as follows. In section two, we set up our notation and solve the KSEs along the
lightcone directions for a class of static M-horizons, which preserve at least one supersymmetry. In sections three and four, we solve the KSEs
for this class of static M-horizons, and investigate the geometry of electric static M-horizons.
We also present several examples. In section five, we solve the KSEs for all static M-horizons, and the field equations
in the electric case. In section six, we give our conclusions.

\newsection{Solution of Killing spinor equations}

\subsection{Static near Horizon Geometry}

To describe the near horizon geometry of 11-dimensional black holes, we shall use
the Gaussian null coordinates of  \cite{gnull} to describe the geometry near the black hole
horizons. In particular assuming appropriate analyticity conditions as well as the existence of an extreme limit and an analysis similar to that done for 5-dimensional supergravity in \cite{reallbh} or for IIB supergravity in \cite{iibh}, we find  that after taking the extreme limit the metric and 4-form field strength of the near horizon geometry of 11-dimensional black holes can be written as
\bea
ds^2 &=& 2 \bbe^+ \bbe^- + \delta_{ij} \bbe^i \bbe^j~,
\cr
F&=& \bbe^+ \wedge \bbe^- \wedge Y
+ r \bbe^+ \wedge d_h Y + X~,~~~~dX=0~,
\eea
where $(u,r, y^I)$ are the coordinates of spacetime, $d_h Y=dY - h \wedge Y$ and
\bea
\label{nhbasish}
\bbe^+ = du~,~~~
\bbe^- = dr + r h - {1 \over 2} r^2 \Delta du~,
~~~\bbe^i =e^i{}_J dy^J~,~~~
\eea
is a frame basis with $h=h_i (y) \bbe^i$ a 1-form and $\Delta=\Delta(y)$  a function
which depend only on the $y$ coordinates. The horizon section ${\cal S}$ is the 9-dimensional submanifold given by $r=u=0$ with metric $ds^2({\cal S})=\delta_{ij} \bbe^i\bbe^j$. Observe
that $\Delta$ and $h$ are a globally defined scalar and 1-form on ${\cal S}$, respectively.

Static horizons are those for which\footnote{We thank James Lucietti for a discussion on this point.}
\bea
\bbe^-\wedge d\bbe^-=0~,
\eea
which yields
\bea
dh=0~,~~~d\Delta=\Delta h~.
\label{stat2}
\eea
Static horizons can be subdivided in two classes. One subclass is to take
\bea
\label{stat2a}
\Delta =0, \qquad dh =0~,
\eea
and other is
\bea
\label{stat2b}
\Delta > 0, \qquad h = d \log \Delta~.
\eea
For supersymmetric horizons $\Delta\geq 0$ since $\partial_u$ is either null or time-like.

In the former case, on introducing a local co-ordinate $x$ such that $h=dx$ and making
a change of co-ordinates $r \rightarrow e^x r$, the metric can be rewritten as
\bea
ds^2 = 2 e^{-x} du dr + ds^2({\cal S})~,
\eea
and the near horizon geometry is a warped product $\bR^{1,1} \times_w {\cal S}$.
However if $h$ is closed but not exact, the resulting warped product is local.

{}For the latter case, the
metric can be rewritten, after a change of coordinates $r\rightarrow r \Delta$, as
\bea
ds^2=2\Delta^{-1} du (dr-{1\over2} r^2 du)+ ds^2({\cal S})~,
\eea
and so the near horizon geometry is a warped product $AdS_2\times_w {\cal S}$.

In the investigation of field and KSEs, it is instructive to begin with backgrounds for which
\bea
h=0~,~~~
\la{hd}
 \eea
 and $\Delta$ an arbitrary function of ${\cal S}$.
It may seem that these horizons are not static because they do not  a priori satisfy the static condition (\ref{stat2}). However, as we shall show the field equations imply that $\Delta$
is constant and so all horizons satisfying (\ref{hd}) are static. The advantage with dealing
with condition (\ref{hd}) is that the solution of the KSEs is particularly simple and the geometry
of the horizons can be easily described.

After understanding the geometry of the (\ref{hd}) horizons, we shall present the solution
of the KSEs for all static horizons without going into details. This is because  static
horizons with $h\not=0$  are more easily investigated in the context of rotating horizons which will be presented elsewhere.

 The metric and 4-form flux  of static $h=0$ horizons become
\bea
ds^2 &=& 2 \bbe^+ \bbe^- + \delta_{ij} \bbe^i \bbe^j~,
\cr
F&=& \bbe^+ \wedge \bbe^- \wedge Y
+ r \bbe^+ \wedge dY + X~,~~~~dX=0~,
\la{snhg}
\eea
where now
\bea
\label{nhbasis}
\bbe^+ = du~,~~~
\bbe^- = dr  - {1 \over 2} r^2 \Delta du~,
~~~\bbe^i =e^i{}_J dy^J~.~~~
\eea
Observe that if $\Delta>0$, the vector field $V=\partial_u$ is time-like and Killing and becomes null at $r=0$ the location of the horizon.  $V$ is identified with the stationary
vector field of the black hole spacetime at the near horizon limit.

We shall consider static near horizon geometries which preserve at least one supersymmetry.
For this, we shall require that (\ref{snhg}) solves the Killing spinor equations
\bea
\nabla_M \epsilon
+\bigg(-{1 \over 288} \Gamma_M{}^{L_1 L_2 L_3 L_4} F_{L_1 L_2 L_3 L_4}
+{1 \over 36} F_{M L_1 L_2 L_3} \Gamma^{L_1 L_2 L_3} \bigg) \epsilon =0~,
\eea
of 11-dimensional supergravity. To achieve this, we shall use spinorial geometry
and the techniques and notation developed in \cite{system11}. In this context, we
set $i,j=1,2,3,4,6,7,8,9,\sharp$, where $\sharp$ is identified with the 10-th
direction,  and the light-cone directions $\bbe^+, \bbe^-$ are spanned by the time and 5-th directions of the spacetime.

We shall also make use of the field equations
\bea
R_{MN} &=& {1 \over 12} F_{M L_1 L_2 L_3} F_N{}^{L_1 L_2 L_3}
-{1 \over 144} g_{MN} F_{L_1 L_2 L_3 L_4}F^{L_1 L_2 L_3 L_4}~,
\cr
d \star  F &=&{1 \over 2} F \wedge F~,
\la{feqns}
\eea
of 11-dimensional supergravity, where the spacetime orientation  is
taken as
\bea
d{\rm vol}_{11}= \bbe^+ \wedge \bbe^- \wedge d{\rm vol}({\cal S}) \ .
\eea
and  $d{\rm vol}({\cal S})=e^{12346789\sharp}$.

\subsection{Light-cone integrability of Killing spinor equations}

The KSEs of 11-dimensional supergravity for the background given in (\ref{snhg}) with $h=0$ and $\Delta$ a function on 
${\cal S}$ can be integrated along the light-cone directions. For this, we decompose the Killing spinor
as
\be
\epsilon = \epsilon_+ + \epsilon_-~,~~~\Gamma_\pm\epsilon_\pm=0~.
\la{spn}
\ee
Then a straightforward calculation reveals that
\bea
\label{ksp1}
\epsilon_+ = \eta_+, \qquad \epsilon_- = \eta_- + r \Gamma_-\Theta_+
\eta_+
\eea
and
\bea
\label{ksp2}
\eta_+ = \phi_+ + u \Gamma_+ \Theta_- \phi_- , \qquad \eta_- = \phi_-
\eea
where now the spinors $\phi_\pm = \phi_\pm (y)$ do not depend on $r$ or $u$, and  we have set
\bea
\Theta_\pm=
{1 \over 288} X_{\ell_1 \ell_2 \ell_3 \ell_4}
\Gamma^{\ell_1 \ell_2 \ell_3 \ell_4} \pm{1 \over 12} Y_{\ell_1 \ell_2} \Gamma^{\ell_1 \ell_2}~.
\eea
In addition, the $+$ and $-$
components of the KSEs impose the following algebraic conditions
\bea
\label{cc1}
&&\big( {1 \over 2} \Delta  +{1 \over 72} dY_{\ell_1 \ell_2 \ell_3}
\Gamma^{\ell_1 \ell_2 \ell_3}
-2 \Theta_-
\Theta_+ \big) \phi_+ =0~,
\eea

\bea
\label{cc2}
&& \bigg(  \partial_i \Delta \Gamma^i
+  {1 \over 6} dY_{\ell_1 \ell_2 \ell_3} \Gamma^{\ell_1 \ell_2
\ell_3}
 \Theta_+ \bigg) \phi_+ =0~,
\eea

\bea
\label{cc3}
&&\bigg( {1 \over 2} \Delta  -{1 \over 72} dY_{\ell_1 \ell_2 \ell_3}
\Gamma^{\ell_1 \ell_2 \ell_3}
-2 \Theta_-
 \Theta_+ \bigg)
 \Theta_-
\phi_- =0~,
 \eea

\bea
\label{cc4}
&& \bigg( {1 \over 4} \partial_i \Delta \Gamma^i
+  {1 \over 24} dY_{\ell_1 \ell_2 \ell_3} \Gamma^{\ell_1 \ell_2
\ell_3}
 \Theta_+ \bigg)
   \Theta_-
\phi_- =0~,
 \eea

\bea
\label{cc5}
&& \bigg( -{1 \over 2} \Delta  +{1 \over 24}
dY_{\ell_1 \ell_2 \ell_3} \Gamma^{\ell_1 \ell_2 \ell_3}
+ 2 \Theta_+
  \Theta_- \bigg)
\phi_- =0~.
 \eea
These further constrain both the spinors $\phi_\pm$, the fluxes and geometry of the horizon.
We shall solve all these conditions as well as the remaining KSEs along the directions of the
horizon section ${\cal S}$ for one Killing spinor.

\newsection{N=1 Supersymmetry}

\subsection{Killing vector bilinear}

Let us assume that (\ref{snhg}) with $h=0$ and $\Delta$ a function of ${\cal S}$  admits one Killing spinor. The existence of a Killing spinor
implies that the spacetime admits a Killing vector field $W$, which is constructed as a spinor bilinear, that is either time-like or null.
The analysis of the KSEs proceeds by identifying the Killing vector bilinear $W$ with the
Killing vector field of the black hole horizon $V$. The associated 1-form of the latter
is given by
\bea
V=\bbe^--{1\over2} r^2\Delta \bbe^+
\eea
while the 1-form associated with the former can be computed
from the expression
\bea
W=\langle B \epsilon^*, \Gamma_M \epsilon \rangle\,\, e^M~,
\la{inner}
\eea
for the 1-form spinor bilinear.

To compute $W$ and compare it to $V$, we have to evaluate the expression for
$W$. For this we shall use the residual  $Spin(9)$ gauge symmetry of KSEs which fixes the
two light-cone directions that we have integrated over.   To proceed
the 32-dimensional Majorana  representation of $Spin(10,1)$ decomposes under $Spin(9)$
into two 16-dimensional Majorana representations. This decomposition has already
been given in (\ref{spn}), where  the Killing spinor was written as a sum of two spinors
with opposite chirality along the light-cone directions. In addition, $Spin(9)$ acts transitively on the $S^{15}$ sphere in
 the 16-dimensional Majorana representation with isotropy group $Spin(7)$, $Spin(9)/Spin(7)=S^{15}$. Using this, the spinor $\phi_-$ can be chosen to lie in any direction and in particular one can set
 \be
\phi_- = w (e_5+e_{12345})~,
\ee
for some real function $w$.
Next, on comparing $W$ and $V$ in the basis ({\ref{nhbasis}}) one finds that
\be
W_+|_{r=0} =0~.
\ee
The $W_+$ component can be computed using (\ref{inner}) and $\phi_- = w (e_5+e_{12345})$ to reveal that  $w=0$. Thus, we find that
\be
\phi_-=0~.
\la{phi-}
\ee

To continue, one can again  use $r,u$-independent $Spin(9)$ transformations  to set, without loss of generality,
\be
\phi_+ = z (1+e_{1234})~,
\ee
for some real ($r,u$-independent) function $z$.  Using this and (\ref{inner}), one finds that
\be
W_- = -2 \sqrt{2} z^2~.
\ee
But $V_-=1$, and so  $z$ is constant. For convenience, we set $z=1$, and so
\be
\phi_+ = 1+e_{1234}~.
\la{phi+}
\ee
To summarize the results so far, substituting (\ref{phi-}) and (\ref{phi+}) into the expression for the Killing spinor $\epsilon$, ({\ref{ksp1}}) and ({\ref{ksp2}}), we find that
\be
\epsilon = (1+e_{1234}) +r \Gamma_- \Theta_+(1+e_{1234})~.
\ee

The Killing spinor epsilon $\epsilon$ can be further simplified. As $\Gamma_+ \Theta_+(1+e_{1234})=0$, $\Theta_+(1+e_{1234})$ is also a $Spin(9)$ Majorana spinor and so it can be expanded in
the basis $1, e_{1234}, e_i, e_{ij}, e_{ijk}$ for $i, j,k=1, \dots , 4$.
Using this and the above expression for the Killing spinor $\epsilon$,  it is straightforward to evaluate the
remaining components of the spinor bilinear $W$ in the directions transverse to the light cone directions,
and determine the resulting constraints imposed on the components of $\Theta_+(1+e_{1234})$.
In particular, requiring that $W_\sharp=0$ implies that the component of $\Theta_+(1+e_{1234})$ in the
$1+e_{1234}$ direction vanishes.

Furthermore, requiring that $W_\alpha=0$
forces the components of $\Theta_+(1+e_{1234})$ in the $e_i$ and $e_{ijk}$ directions to vanish as well. As a result $\Theta_+(1+e_{1234})$ must be a linear combination of $i(1-e_{1234})$ and $e_{ij}$ and so must lie in the vector representation of $Spin(7)$ the isotropy group of $1+e_{1234}$. On the other hand $Spin(7)$ acts transitively on the $S^6$ sphere in the 7-dimensional vector representation with isotropy group $SU(4)$. As a result
$\Theta_+(1+e_{1234})$ can be chosen to lie in any direction and in particular
one can then without loss of generality take
\be
\Theta_+(1+e_{1234})= i \Phi (1-e_{1234})~,
\ee
for some real function $\Phi=\Phi(y)$. On examining the component $W_+$ of the Killing spinor bilinear, one finds that
\be
\Delta = 4 \Phi^2~.
\label{fv6}
\ee
This concludes all the conditions on the Killing spinor which arise from the identification
of $W$ with the Killing vector field of the black hole horizon.

After considering the  KSEs along the directions transverse to the light-cone, the independent
conditions which have to be solved so that the near horizon geometry (\ref{snhg})  admits
at least one supersymmetry are

\bea
\label{fv3}
  \Theta_+(1+e_{1234}) = i \Phi (1-e_{1234})~,
\eea
\bea
\label{fv4}
\hn_i (1+e_{1234}) + \bigg({1 \over 24} X_{i \ell_1 \ell_2 \ell_3} \Gamma^{\ell_1 \ell_2 \ell_3}
+{1 \over 8} \Gamma_i{}^{\ell_1 \ell_2} Y_{\ell_1 \ell_2} \bigg) (1+e_{1234})
\nn
-i \Phi \Gamma_i (1-e_{1234})=0~,
\eea
\bea
\label{fv5}
\hn_i (i \Phi (1-e_{1234})) + i \Phi \bigg(
-{1 \over 24} X_{i \ell_1 \ell_2 \ell_3} \Gamma^{\ell_1 \ell_2 \ell_3}
+{1 \over 8} \Gamma_i{}^{\ell_1 \ell_2} Y_{\ell_1 \ell_2} \bigg) (1-e_{1234})
\nn
+ \bigg( {1 \over 4} \Delta \Gamma_i  -{1 \over 48} dY_{\ell_1 \ell_2 \ell_3}
\Gamma^{\ell_1 \ell_2 \ell_3} \Gamma_i \bigg) (1+e_{1234})=0~,
\eea
where $\Delta$ and $\Phi$ are related as in (\ref{fv6}), and $\hn$ is the Levi-Civita
connection on the horizon section ${\cal S}$.
We remark that conditions ({\ref{cc1}}) and ({\ref{cc2}}) have been omitted from this list, because they are implied by ({\ref{fv3}}), ({\ref{fv4}}),
({\ref{fv5}}) and ({\ref{fv6}}).

\subsection{Solution to the Killing spinor equations}

The KSEs (\ref{fv3})-(\ref{fv5}) can be easily solved using the spinorial geometry techniques of \cite{spingem, system11} and the general results of \cite{system11}. In particular, the differential
and algebraic conditions turn into a linear system for the geometry as expressed in terms of the spin connection and the components of the fluxes. This system can be solved to express
some of the components of the fluxes in terms of the geometry and find the conditions on the
spacetime geometry imposed by supersymmetry.

Before we proceed with the solution to the linear system, the spacetime admits
an 1-form, 2-form and 5-form bilinear. As a consequence of the assumption that the Killing spinors are globally defined, all these three bilinears are also globally defined on the spacetime. We have already stated the 1-form bilinear. The remaining  two are
\bea
\alpha=2 (\bbe^-+{1\over2} r^2 \Delta \bbe^+)\wedge e^\sharp-4 r \Phi \omega~,
\eea
and
\bea
\sigma=- \{(\bbe^-+{1\over2} \Delta r^2 \bbe^++2i r\Phi \bbe^\sharp)\wedge \chi+ {\rm c.c.}\}+ (\bbe^--{1\over2} \Delta r^2 \bbe^+)\wedge
\omega\wedge \omega~,
\eea
where $\omega=-i\delta_{\a\bar\b} e^\a\wedge e^{\bar\b}$ is a Hermitian 2-form and $\chi$ is a (4,0)-form
on the directions transverse to the light-cone and $e^\sharp$. Taking the light-cone
directions as globally defined, $e^\sharp$, $\omega$ and $\chi$ are also globally defined.
Note that the index $i$ transverse to the light-cone directions decomposes as $i=\a, \bar\a, \sharp$, where $\a=1,2,3,4$. As a result both the near horizon geometry and the horizon section ${\cal S}$ admit an
$SU(4)$ structure.

We shall not give the linear system as it is easily derived from the KSEs. The solution
of the linear system expresses the flux $Y$ in terms of the geometry as
\bea
Y=-de^\sharp-2 \Phi\,\omega~,~~~
\la{q1}
\eea
and $\Phi$  as
\bea
\Phi=-{i\over2} \big( \Omega_{\sharp, \lambda}{}^\lambda-\Omega_{\lambda, \sharp}{}^\lambda \big)~,
\eea
respectively.
Also, $\Phi$ satisfies
\be
\label{q5}
\partial_\alpha \Phi = \Phi \big(
-2  \Omega_{\bar{\lambda},}{}^{\bar{\lambda}}{}_\alpha  +2 \Omega_{\alpha,\lambda}{}^\lambda
- \Omega_{\sharp, \sharp \alpha} \big)~,
\ee
\be
\label{q6}
\partial_\sharp \Phi = -{1 \over 4} \Phi
X_\lambda{}^\lambda{}_\sigma{}^\sigma~.
\ee
The 4-form $X$ is expressed in terms of the geometry as
\be
\label{q7}
X_{\mu \bar{\lambda}_1 \bar{\lambda}_2 \bar{\lambda}_3}
= \bigg(-\Omega_{\mu, \sharp \sigma}+\Omega_{\sharp, \mu \sigma}
-2 \Omega_{[\mu|, \sharp |\sigma]}
 \bigg) \epsilon^\sigma{}_{ \bar{\lambda}_1 \bar{\lambda}_2 \bar{\lambda}_3}~,
\ee
\be
\label{q8}
X_{\bar{\beta} \alpha \lambda}{}^\lambda + {1 \over 4}  X_\lambda{}^\lambda{}_\sigma{}^\sigma \delta_{\bar{\beta} \alpha}
= \Omega_{\bar{\beta}, \sharp \alpha}+ \Omega_{\alpha, \sharp \bar{\beta}}~,
\ee
\be
\label{q9}
{2 \over 3} \Omega_{\sharp, \lambda}{}^\lambda
+{2 \over 3} \Omega_{\lambda, \sharp}{}^\lambda +{1 \over 6}X_\lambda{}^\lambda{}_\sigma{}^\sigma
+{1 \over 18} X_{\lambda_1 \lambda_2 \lambda_3 \lambda_4} \epsilon^{\lambda_1 \lambda_2 \lambda_3 \lambda_4}=0~,
\ee
\be
\label{q10}
X_{\sharp \lambda_1 \lambda_2 \lambda_3} =
 \big(\Omega_{\bar{\sigma}, \lambda}{}^\lambda
-{1 \over 2}\Omega_{\sharp, \sharp \bar{\sigma}} \big) \epsilon^{\bar{\sigma}}{}_{\lambda_1 \lambda_2 \lambda_3}~,
\ee
\bea
\label{q11}
X_{\sharp \beta \bar{\sigma}_1 \bar{\sigma}_2}
&=& {2 \over 3} \big(\Omega_{\beta, \mu_1 \mu_2}
+ \Omega_{\mu_1, \beta \mu_2} \big) \epsilon^{\mu_1 \mu_2}{}_{\bar{\sigma}_1 \bar{\sigma}_2}
-2 \Omega_{\beta, \bar{\sigma}_1 \bar{\sigma}_2}
\nn
&+& \bigg(-{4 \over 3} \Omega_{\lambda ,}{}^\lambda{}_{[ {\bar{\sigma}_1}}
+{4 \over 3} \Omega_{[\bar{\sigma}_1 , |\lambda|}{}^\lambda
+{2 \over 3} \Omega_{\sharp, \sharp [{\bar{\sigma}}_1} \bigg) \delta_{\bar{\sigma}_2 ] \beta}~.
\eea
The conditions on the geometry are
\bea
\label{q12}
\Omega^\lambda{}_{,\sharp \lambda} +\Omega_{\lambda, \sharp}{}^\lambda=0~,
\cr
 -2  \Omega_{\bar{\lambda}_1 , \bar{\lambda}_2 \bar{\lambda}_3}
\epsilon^{\bar{\lambda}_1 \bar{\lambda}_2 \bar{\lambda}_3}{}_\alpha +4
\Omega_{\bar{\lambda},}{}^{\bar{\lambda}}{}_\alpha -2 \Omega_{\alpha, \lambda}{}^\lambda
+ \Omega_{\sharp, \sharp \alpha}=0~,
\cr
\Omega_{[\mu_1, |\sharp| \mu_2]}-\Omega_{\sharp, \mu_1 \mu_2}
-{1 \over 2} \big(\Omega_{\bar{\sigma}_1, \sharp \bar{\sigma}_2}
-\Omega_{\sharp, \bar{\sigma}_1 \bar{\sigma}_2} \big) \epsilon^{\bar{\sigma}_1 \bar{\sigma}_2}{}_{\mu_1 \mu_2}=0~.
\eea
Observe that the (2,2) and traceless part of $X$ is not determined by the KSEs in terms
of the geometry.

\subsection{Field equations}

In addition to the KSEs the near horizon geometries must satisfy the field
equations (\ref{feqns})
In particular, $M=+, N=+$ component of the Einstein equation gives
\be
{1 \over 2} {\tilde{\nabla}}^2 \Delta = {1 \over 12} (dY)_{\ell_1 \ell_2 \ell_3} (dY)^{\ell_1 \ell_2 \ell_3}~.
\ee
As ${\cal S}$ is assumed to be compact, this implies that $\Delta$ is constant, and
\be
dY=0~.
\ee
Also note that the $M=+, N=-$ component of the Einstein equation gives
\be
\label{delteqn}
\Delta = {1 \over 6} Y_{\ell_1 \ell_2} Y^{\ell_1 \ell_2} +{1 \over 144} X_{\ell_1 \ell_2 \ell_3 \ell_4}
X^{\ell_1 \ell_2 \ell_3 \ell_4}~.
\la{dell}
\ee
If $\Delta=0$ then this condition implies that $Y=0$ and $X=0$, and hence the 4-form also vanishes;
in this case the spacetime is $\bR^{1,1} \times {\cal{S}}$, where ${\cal{S}}$ is a compact $Spin(7)$
holonomy manifold.

For solutions with $\Delta \neq 0$, as $\Delta$ is constant, the  near horizon geometry
is a product $AdS_2\times {\cal S}$.
Since $\Phi$ is constant, ({\ref{q1}}) and $dY=0$ imply that
\be
d {{\omega}}=0~.
\ee
Hence one finds the following additional conditions on the spin connection
\be
\Omega_{[\alpha_1, \alpha_2 \alpha_3]}=0, \qquad \Omega_{\bar{\alpha}, \beta_1 \beta_2}=0,
\qquad - \Omega_{\sharp, \mu_1 \mu_2}+\Omega_{[\mu_1, |\sharp|\mu_2]}=0, \qquad
\Omega_{(\alpha, \bar{\beta}) \sharp}=0~.
\la{gem1}
\ee
Comparing these to the geometric conditions derived from the KSEs (\ref{q12}) and (\ref{q5}), one finds the additional  remaining geometric condition
\be
2 \Omega_{\alpha, \beta}{}^\beta - \Omega_{\sharp, \sharp \alpha} =0~.
\la{gem2}
\ee
Implementing all the geometric conditions (\ref{gem1}) and (\ref{gem2}) on the fluxes, we
find
\bea
&& X_{\sharp\a_1\a_2\a_3}=-\Omega_{\sharp, \sharp\bar\b} \epsilon^{\bar\b}{}_{\a_1\a_2\a_3}~,~~~X_{\sharp\a\bar\b_1\bar\b_2}=\Omega_{\a, \g_1\g_2} \epsilon^{\g_1\g_2}{}_{\bar\b_1\bar\b_2}~,~~~X_{\a\bar\b\g}{}^\g=0~,
\cr
&& X_{\a\bar\b_1\bar\b_2\bar\b_3}=-(\Omega_{\a,\sharp \g}+\Omega_{\sharp, \a\g} ) \epsilon^\g{}_{\bar\b_1\bar\b_2\bar\b_3}~,~~~X_{\b_1\b_2\b_3\b_4}=-{1\over2} (\Omega_{\sharp, \a}{}^\a+\Omega_{\a, \sharp}{}^\a) \epsilon_{\b_1\b_2\b_3\b_4}~,
\cr
&&\Phi = -{i \over 2} \big( \Omega_{\sharp, \lambda}{}^\lambda
- \Omega_{\lambda, \sharp}{}^\lambda \big)~.
\eea
The geometric conditions that we have found (\ref{gem1}) and (\ref{gem2}), as well as the expression
for the fluxes, appear to be expressed in a non-covariant form. However this is not the case. The
components of the spin connection that appear can be naturally identified with
representations of the structure group $SU(4)$ and are related to intrinsic torsion. So they
transform covariantly under the gluing transformations of the tangent bundle which take values in the
structure group. Furthermore observe that $e^\sharp$ is not a Killing direction. However if we take it to
be Killing, then ${\cal S}$ is a fibration over an 8-dimensional almost Hermitian symplectic manifold $B$. The connection of the fibration is proportional to that of the
canonical  bundle of $B$.

Further progress depends on the use of compactness of ${\cal S}$ to solve the remaining
field equations. In fact, it suffices to solve the field equations of the 3-form
gauge potential as these imply all the remaining Einstein equations \cite{system11}. However progress towards this goal is hampered because not all components of $X$ are expressed in terms of the geometry. Some additional difficulty is also encountered  because  the geometric conditions found so far do not imply that the 1-form $e^\sharp$
is associated with a Killing vector field.  As a consequence from now on, we shall focus
on the electric case where $X=0$.

\newsection{Electric static $h=0$ horizons}

\subsection{Solution of field and Killing spinor equations}

Implementing the requirement that $X=0$ (and $\Delta \neq 0$) for electric static horizons and taking into account
the geometric conditions (\ref{gem1}) and (\ref{gem2}), one finds that the geometry
of ${\cal S}$ satisfies
\bea
&&\Omega_{\alpha_1, \alpha_2 \alpha_3}=0, \qquad \Omega_{\bar{\alpha}, \beta_1 \beta_2}=0,
\qquad \Omega_{\sharp, \mu_1 \mu_2}=\Omega_{\mu_1, \sharp\mu_2}=0,
\cr
&&
\Omega_{(\alpha, \bar{\beta}) \sharp}=0~,~~~\Omega_{\sharp, \sharp\a}=0~,~~~
\Phi = -i  \Omega_{\sharp, \a}{}^\a
=i \Omega_{\a, \sharp}{}^\a ~,~~~\Omega_{\alpha, \beta}{}^\beta=0~.
\la{gem3}
\eea
Moreover, $Y=-de^\sharp-2\Phi\omega$ and it is constrained as
\bea
\Delta = {1 \over 6} Y_{\ell_1 \ell_2} Y^{\ell_1 \ell_2}~.
\eea
In addition, the 3-form flux field equations imply that $Y$ is  co-closed on ${\cal S}$. As $dY=0$, $Y$ is harmonic.
We remark that these conditions are sufficient to ensure that the solution preserves (at least) $N=2$ supersymmetry.
To see this, note that ({\ref{fv3}}), ({\ref{fv4}}) and ({\ref{fv5}}) are also satisfied if one replaces the
Majorana spinors $1+e_{1234}$ and $i(1-e_{1234})$ by $i(1-e_{1234})$ and $-(1+e_{1234})$ respectively throughout.

First observe that the geometric conditions (\ref{gem3}) imply that the vector field associated to $e^\sharp$ is Killing on the horizon section ${\cal S}$
and of constant length. Therefore the metric on ${\cal S}$ can be written as
\bea
ds^2({\cal S})= (d\tau+ \lambda)^2+ ds^2(B)
\eea
where $\tau$ is the coordinate along the Killing vector field, $\lambda$ is a 1-form
on the base space $B$. Thus ${\cal S}$ can be thought of as a $U(1)$ fibration over a 8-dimensional manifold $B$.
Furthermore, $i_\sharp\omega=0$ and $\omega$ is invariant under the action of the $e^\sharp$
vector field, and so it descends to a closed (almost) Hermitian form on $B$. Since in addition $B$ is complex, one concludes that $B$ is K\"ahler.
The geometric conditions also imply that curvature of the fibration $de^\sharp$ is (1,1)
and its trace is constant. As a result $e^\sharp$  is a Hermitian-Einstein connection with a non-vanishing cosmological constant $\Phi$.

These restrictions on the fibration solve all the conditions in (\ref{gem3}) apart from
\bea
\Omega_{\alpha, \beta}{}^\beta=0~,~~~\Phi = -i  \Omega_{\sharp, \lambda}{}^\lambda~,~~~\Delta = {1 \over 6} Y_{\ell_1 \ell_2} Y^{\ell_1 \ell_2}~~.
\la{gem4}
\eea
It is clear that the first two conditions can be expressed in terms of components
of $d\chi$. In particular the first condition imposes a certain restriction on the
canonical class of $B$. However, the identification of the precise condition is not apparent
as  $\Omega_{\sharp, \lambda}{}^\lambda\not=0$ which indicates that the chosen frame
$e^\a$ depends on the coordinate along $e^\sharp$, even though the metric and $\omega$ do not, and so it is not adapted to the fibration. We shall illustrate this with an example.

\subsection{Example}

We shall demonstrate that $AdS_2\times S^3\times CY_6$ is a solution, where $CY_6$ is any Calabi-Yau 6-dimensional manifold. In such case ${\cal S}=S^3\times CY_6$. To see this, parameterize $S^3$ in terms of Euler angles
as
\bea
\sigma^1&=& \sin{\psi\over2} \sin\theta d\phi+\cos{\psi\over2} d\theta~,~~~
\sigma^2=-\cos{\psi\over2} \sin\theta d\phi+\sin{\psi\over2} d\theta~,
\cr
\sigma^3&=&{1\over2} d\psi+\cos\theta d\phi~.
\eea
Then write the metric on ${\cal S}$ as
\bea
ds^2({\cal S})= (e^\sharp)^2+ 2 e^1 e^{\bar 1}+ 2\sum_{\a, \b>1} \delta_{\a\bar\b} e^\a e^{\bar\b}~,
\eea
where
\bea
e^\sharp=\sigma^3~,~~~e^1={\sigma^1+i\sigma^2\over\sqrt{2}}={1\over\sqrt{2}} e^{{i\over2}\psi} (-i \sin\theta d\phi+ d\theta)~,~
\eea
and $e^\a$, $\a>1$, a frame on $CY_6$ which is independent from the coordinates of $S^3$.
Observe that $e^1$ depends explicitly on the coordinate $\psi$ of the isometry but
neither the metric nor the K\"ahler form
\bea
\omega=-i e^1\wedge e^{\bar 1}+\omega_{(6)}~,~~~\omega_{(6)}=-i\sum_{\a, \b>1} \delta_{\a\bar\b}\, e^\a\wedge e^{\bar\b}~,
\eea
on  the base space $B=S^2\times CY_6$, depend on $\psi$, where $\omega_{(6)}$ is the K\"ahler form on $CY_6$.
This gives
\bea
\Omega_{\sharp, 1}{}^1={i\over2}~,~~~\Omega_{\sharp, \a}{}^\a=0~~~~{\mathrm {for}}~~~\a>1~,
\eea
and so $\Phi={1\over2}$.

Moreover
\bea
\Omega_{1,1\bar1}=0~,~~~
\eea
and so
\bea
\Omega_{\a,\b}{}^\b=0
\eea

Observe also that
\bea
Y=-\omega_{(6)}
\eea
and so $\Delta = {1 \over 6} Y_{\ell_1 \ell_2} Y^{\ell_1 \ell_2}$.

\subsection{General electric static horizons}

Since the metric and K\"ahler form  $\omega$ are independent from the coordinate $\tau$ along the
isometry, there is a $\tau$-independent frame ${\mathring e}^\a$ and a unitary transformation
$U$, which may depend on all coordinates of ${\cal S}$, such that
\bea
e^\a=U^\a{}_\b {\mathring e}^\b~.
\eea
However $e^\a$ is defined up to a local $SU(4)$ transformation which preserves the Killing spinor $\epsilon$ and so all the conditions we have derived from the field and KSEs. As a result, such a transformation can be used to specify $U$ up to a phase. Thus we can write
\bea
e^\a=e^{i \xi} {\mathring e}^\b~,
\eea
where $\xi$ can depend on all coordinates of ${\cal S}$. As a result
\bea
\Omega_I{}^\a{}_\b=-e^{-i\xi} \partial_I e^{i\xi} \delta^\a{}_\b+ {\mathring\Omega}_I{}^\a{}_\b~,
\eea
where ${\mathring\Omega}$ is the spin connection of the frame $(e^\sharp, {\mathring e}^\a, {\mathring e}^{\bar\a})$, with  $({\mathring e}^\a, {\mathring e}^{\bar\a})$ adapted to $B$. To specify the geometry of $B$, we have to determine the restrictions
on ${\mathring\Omega}$  implied by (\ref{gem4}). In particular
\bea
\Omega_{\sharp, \a}{}^\a=4 e^{-i\xi} \partial_\sharp e^{i\xi} +   {\mathring\Omega}_{\sharp, \a}{}^\a  =i\Phi \ ,
\qquad  {\mathring\Omega}_{\sharp, \a}{}^\a = -i \Phi
\eea
and so
\bea
\partial_\tau\xi={1\over 2}\Phi
\eea leading to $\xi={1\over2}\Phi\, \tau+ \b$, where $\b$ does not
dependent on $\tau$. The gauge transformation generated by $\beta$
is inconsequential as it can be absorbed in the definition of the
frame ${\mathring e}^\a$. So without loss of generality, we can set
\bea \xi={1\over2}\Phi\, \tau~. \eea

Next, we have
\bea
\Omega_{\b,\a}{}^\a=4e^{-i\xi} \partial_\b e^{i\xi}+{\mathring\Omega}_{\b,\a}{}^\a=0 \ .
\eea
Thus, we find
\bea
{\mathring\Omega}_{\b,\a}{}^\a=2i \Phi \lambda_\b~.
\eea
So the Ricci form of $B$ is
\bea
{\mathring \rho}\equiv -i {\mathring R}_{\b\bar\g,\a}{}^\a {\mathring e}^\b\wedge {\mathring e}^{\bar\g}= 2\Phi\, de^\sharp~.
\eea
Clearly the curvature of the canonical bundle of $B$ is proportional to that the
fibration of ${\cal S}$ over $B$.
In terms of the Ricci form, the last condition in (\ref{gem4}) can be rewritten as
\bea
{1\over6} \hat{\mathring\rho}_{ij} \hat{\mathring\rho}^{ij}=4\Phi^4~,
\eea
where
\bea
\hat{\mathring\rho}_{ij}={\mathring\rho}_{ij}+ \Phi^2\, \omega_{ij}~,~~~\mathring\rho_\a{}^\a=4i\Phi^2
\eea
is the (1,1) and traceless component of the curvature of the canonical bundle. Thus $B$
is a K\"ahler manifold for which the canonical bundle is equipped with a Hermitian-Einstein
connection, the Ricci curvature has point-wise constant length but it is not Einstein, and the Ricci scalar is constant. A consequence of the Hermitian-Einstein condition for the connection of the canonical bundle is that  the K\"ahler metric on $B$ is Yamabe, ie the Ricci scalar is constant.

\subsection{More examples}

To classify the electric static horizons, one has to find the 8-dimensional
K\"ahler-Yamabe manifolds which admit a K\"ahler metric  such that
the Ricci tensor has point-wise constant length.  To find examples, we shall take $B$ to be a product of K\"ahler-Einstein and Calabi-Yau manifolds $N_p$ of real dimension $2 n_p$. Such examples have also been constructed in \cite{kim, kim2} in the context of AdS${}_2$/CFT${}_1$ correspondence. So the Ricci forms are
\bea
\rho_p= \Phi \ell_p \omega_p~,~~~p\leq 4~,~~~\sum_p n_p=4~,
\eea
with
\bea
{\mathring \rho}=\sum_p \rho_p~,~~~\omega=\sum_p \omega_p~,
\eea
where $\omega_p$ is the K\"ahler form of $N_p$.  The geometric
conditions imply that
\bea
\sum_p n_p \ell_p=-4\Phi~,~~~\sum_p n_p \ell_p^2=16 \Phi^2~.
\eea
Solutions to these equations will give examples of near horizon geometries. For the explicit example given above
$B=S^2\times CY_6$ and so $n_1=1, n_2=3$ and $\ell_1=-4\Phi, \ell_2=0$. For more examples, take
$B=N_1\times N_2$, where $N_1, N_2$ are 4-dimensional K\"ahler Einstein spaces and so $n_1=n_2=2$.
 The conditions on the geometry imply that
\bea
\ell_1=(-1\pm\sqrt 3)\Phi~,~~~\ell_2=(-1\mp\sqrt 3)\Phi~.
\eea
In either case, one of the spaces has negative Ricci curvature. Since all K\"ahler manifolds
with negative first Chern class admit Einstein metrics and there are 4-dimensional K\"ahler
manifolds with positive first Chern class admitting Einstein metrics, there are many examples
of  electric static horizons.

\newsection{Static Horizons}

Now we shall turn to the investigation of static horizons with $h\not=0$. There are two classes described by the  conditions (\ref{stat2a}) and (\ref{stat2b}), respectively.  The  $M=+, N=+$ component of the Einstein equation can be written as
\bea
{1 \over 2} {\tilde{\nabla}}^2 \Delta -{3 \over 2} h^i {\tilde{\nabla}}_i \Delta
-{1 \over 2} \Delta {\tilde{\nabla}}^i h_i + \Delta h^2 = {1 \over 12}
\big( dY - h \wedge Y \big)_{ijk} \big( dY - h \wedge Y \big)^{ijk}
\eea
For both  ({\ref{stat2a}}) and ({\ref{stat2b}}) cases, the LHS of this equation vanishes
identically, and we therefore find that
\be
dY - h \wedge Y =0~.
\ee
From now on, we shall investigate the two classes separately.

\subsection{Static $dh=\Delta=0$ horizons}

\subsubsection{Solution of KSEs}

As we have already mentioned, the solution of the KSEs for static horizons with $h\not=0$ are
a special case of that of rotating horizons. Because of this, we shall not explain the solution
of the KSEs in detail. Instead, we shall simply state the solution. It turns out that since $\Delta=0$, the Killing spinor is
\bea
\e=1+e_{1234}~.
\eea
Substituting this and ({\ref{stat2a}}) into the KSEs, one finds the spacetime geometry  is restricted as
\bea
2 \Omega_{\sharp, \lambda}{}^\lambda-\Omega_{\lambda, \sharp}{}^\lambda
+\Omega_{{\bar{\lambda}}, \sharp}{}^{{\bar{\lambda}}}= 0~,
\eea
\bea
\Omega_{\sharp, \mu_1 \mu_2} - \Omega_{[\mu_1, |\sharp| \mu_2]}
-{1 \over 2} \big( \Omega_{\sharp, {\bar{\lambda}}_1 {\bar{\lambda}}_2}
- \Omega_{[{\bar{\lambda}}_1, |\sharp| {\bar{\lambda}}_2]} \big)
\epsilon^{{\bar{\lambda}}_1 {\bar{\lambda}}_2}{}_{\mu_1 \mu_2} =0~,
\eea
\bea
h_\sharp &=& -{1 \over 2} \big(\Omega_{\lambda, \sharp}{}^\lambda
+\Omega_{{\bar{\lambda}}, \sharp}{}^{{\bar{\lambda}}} \big)~,
\nn
h_\alpha &=& -{2 \over 3} \Omega_{{\bar{\lambda}}_1, {\bar{\lambda}}_2 {\bar{\lambda}}_3}
\epsilon^{{\bar{\lambda}}_1 {\bar{\lambda}}_2 {\bar{\lambda}}_3}{}_\alpha
+{4 \over 3} \Omega_{\bar{\beta},}{}^{\bar{\beta}}{}_\alpha
-{2 \over 3} \Omega_{\alpha,\beta}{}^\beta +{1 \over 3} \Omega_{\sharp, \sharp \alpha}~.
\eea
Observe that $h$ is specified in terms of the Levi-Civita connection along the horizon section directions. In addition, one has to impose
\be
dh =0~.
\ee
In addition some of the components of the flux can be expressed in terms of the geometry. In particular one finds that
\bea
Y= - \bbe^\sharp \wedge h -d \bbe^\sharp~,
\eea
and
\bea
{1 \over 3} X_{\mu_1 \mu_2 \mu_3 \mu_4} \epsilon^{\mu_1
\mu_2 \mu_3 \mu_4}+ X_{\sigma}{}^\sigma{}_\rho{}^\rho &=& \Omega_{{\bar{\lambda}},\sharp}{}^{\bar{\lambda}}-7 \Omega_{\lambda, \sharp}{}^\lambda~,
\eea
\bea
X_{\sharp \mu_1 \mu_2 \mu_3} = -2 \Omega_{[\mu_1, \mu_2 \mu_3]}
-{2 \over 3} \big( \Omega_{\nu,}{}^\nu{}_{\bar{\sigma}}
-\Omega_{\bar{\sigma},\nu}{}^\nu +\Omega_{\sharp,\sharp \bar{\sigma}}
\big)\epsilon^{\bar{\sigma}}{}_{\mu_1 \mu_2 \mu_3}~,
\eea
\bea
X_{\sharp \beta {\bar{\sigma}}_1 {\bar{\sigma}}_2}
&=& {2 \over 3} \big(\Omega_{\beta, \mu_1 \mu_2}+ \Omega_{\mu_1, \beta \mu_2} \big)
\epsilon^{\mu_1 \mu_2}{}_{{\bar{\sigma}}_1 {\bar{\sigma}}_2} -2 \Omega_{\beta, {\bar{\sigma}}_1
{\bar{\sigma}}_2}
\nn
&+& \big(-{4 \over 3} \Omega_{\nu,}{}^\nu{}_{[{\bar{\sigma}}_1}+{4 \over 3} \Omega_{[{\bar{\sigma}}_1,}{}_\nu{}^\nu
+{2 \over 3} \Omega_{\sharp, \sharp [{\bar{\sigma}}_1} \big) \delta_{{\bar{\sigma}}_2] \beta}~,
\eea
\bea
X_{\alpha {\bar{\beta}} \lambda}{}^\lambda -{1 \over 4} \delta_{\alpha \bar{\beta}} X_{\sigma}{}^\sigma{}_\rho{}^\rho
= -2 \Omega_{(\alpha, |\sharp| {\bar{\beta}})} +{1 \over 4}\big(\Omega_{\lambda, \sharp}{}^\lambda
+\Omega_{{\bar{\lambda}}, \sharp}{}^{{\bar{\lambda}}} \big) \delta_{\alpha \bar{\beta}}~,
\eea
\bea
X_{\mu {\bar{\sigma}}_1 {\bar{\sigma}}_2 {\bar{\sigma}}_3}
-{1 \over 2} X_{\mu \sigma \lambda}{}^\lambda \epsilon^\sigma{}_{{\bar{\sigma}}_1 {\bar{\sigma}}_2 {\bar{\sigma}}_3}
= - \big(\Omega_{\mu, \sharp \sigma}+\Omega_{[\mu, |\sharp| \sigma]}
-\Omega_{[{\bar{\lambda}}_1, |\sharp| {\bar{\lambda}}_2]} \epsilon^{{\bar{\lambda}}_1
{\bar{\lambda}}_2}{}_{\mu \sigma} \big)  \epsilon^\sigma{}_{{\bar{\sigma}}_1 {\bar{\sigma}}_2 {\bar{\sigma}}_3}~.
\eea
Observe again that the (2,2) and traceless component of the magnetic flux $X$ is not constrained
by the KSEs.

The isotropy subgroup of the Killing spinor in  $Spin(10,1)$ is $Spin(7)\ltimes \bR^9$.
Therefore, the horizon section ${\cal S}$ admits a $Spin(7)$ structure. In particular although we have
decomposed the conditions that arise from the KSEs in $SU(4)$ representations, they can be rewritten in terms
of $Spin(7)$ representations. Further investigation of  the geometry of ${\cal S}$ requires the solution of the
field equations. As in the previous case, this is rather involved in the presence of magnetic fluxes $X$. So we shall set $X=0$
and explore the geometry of electric horizons.

\subsubsection{Electric static $dh=\Delta=0$ horizons}

The  $Y$ flux of electric, $X=0$, static horizons can be rewritten
as
\bea
Y = -{3 \over 2} \bbe^\sharp \wedge h + Z , \qquad
i_{\bbe^\sharp} Z=0~.
\eea
The additional conditions on the geometry
obtained by setting $X=0$ in the expressions for the fluxes in the
previous section  imply \bea h_\sharp =0 \ . \eea Then the $M=+,
N=-$ component of the Einstein equations can be rewritten as
\bea
{\tilde{\nabla}}^i h_i = -{1 \over 4} h^2 -{1 \over 6}
Z_{ij}Z^{ij}~.
\eea
On integrating both sides of this condition over
${\cal{S}}$, one finds that $h=0$ and $Z=0$. So $Y=0$ and since
$X=0$, the  4-form flux $F$ vanishes.  The spacetime is a product
$\bR^{1,1} \times {\cal{S}}$, where ${\cal{S}}$ is a compact
$Spin(7)$ holonomy manifold. The Berger classification in turn
implies that locally ${\cal S}=S^1\times N$, where $N$ is an
8-dimensional holonomy $Spin(7)$ manifold.

\subsection{Static  $h=d\log\Delta$ horizons}

\subsubsection{Solution of KSEs}

Next let us turn to the  solution of the KSEs for static horizons  satisfying ({\ref{stat2b}}). The Killing spinor in this case can be chosen as
\bea
\e=1+e_{1234}+i r \Phi \Gamma_- (1-e_{1234})~,
\eea
where $\Delta=4\Phi^2$. In fact $\Phi$ can be chosen to be a positive function\footnote{Changing the sign of $\Phi$ corresponds to a sign choice for the
almost complex structure on the horizon section.} up to a $Spin(7)$ gauge transformation.
Substituting this and (\ref{stat2b}) into the KSEs, one finds the conditions
\bea
d \big(\Delta^{-{1 \over 2}} \omega \big) =0~,
\la{wg1}
\eea
\bea
-2 \Omega_{\alpha, \beta}{}^\beta + \Omega_{\sharp, \sharp \alpha} =0~,
\la{wg2}
\eea
\bea
 \Delta^{1 \over 2} = -i \big( \Omega_{\sharp, \lambda}{}^\lambda +{1 \over 2} \Omega_{\bar{\lambda}, \sharp}
{}^{\bar{\lambda}} - {1 \over 2}\Omega_{\lambda, \sharp}{}^\lambda \big)~,
\la{wg3}
\eea
on the geometry of spacetime, where $\omega=-i\delta_{\a\bar\b} e^\a\wedge e^{\bar\b}$.

In addition, the KSEs express some of the fluxes in terms of the geometry as
\bea
Y = -d \bbe^\sharp - \Delta^{{1 \over 2}} \omega - \Delta^{-1} \bbe^\sharp \wedge d \Delta
\eea
and
\bea
X_{\sharp\a_1\a_2\a_3}&=&\big(-\Omega_{\sharp, \sharp\bar\b}
-{1 \over 2} \Delta^{-1} \partial_{\bar\b} \Delta \big) \epsilon^{\bar\b}{}_{\a_1\a_2\a_3}~,~~~
X_{\sharp\a\bar\b_1\bar\b_2}=\Omega_{\a, \g_1\g_2} \epsilon^{\g_1\g_2}{}_{\bar\b_1\bar\b_2}~,
\nn
X_{\a\bar\b\g}{}^\g &=&0~,~~~
X_{\a\bar\b_1\bar\b_2\bar\b_3}=-(\Omega_{\a,\sharp \g}+\Omega_{\sharp, \a\g} ) \epsilon^\g{}_{\bar\b_1\bar\b_2\bar\b_3}~,
\nn
X_{\b_1\b_2\b_3\b_4}&=&-{1\over2} (\Omega_{\sharp, \a}{}^\a+\Omega_{\a, \sharp}{}^\a -{1 \over 2}
\Delta^{-1} \partial_\sharp \Delta) \epsilon_{\b_1\b_2\b_3\b_4}~.
\eea
Observe again that the (2,2) and traceless part of the $X$ is not determined in terms of the
geometry.

The spacetime is a warped product $AdS_2\times_w {\cal S}$, where ${\cal S}$ admits
a $SU(4)$ structure. The $SU(4)$ structure is further restricted by the geometric conditions (\ref{wg1})-(\ref{wg3}). Although ${\cal S}$ admits a preferred direction $\bbe^\sharp$,
generically this direction is not an isometry. Moreover, the almost complex structure
in the 8-dimensions transverse to $\bbe^\sharp$ is not integrable. However, ${\cal S}$
admits a conformally symplectic form in the directions transverse to $\bbe^\sharp$.

To proceed, it is convenient to introduce a new frame  $\hat\bbe$ on ${\cal S}$
 as
\bea
\label{conf2}
\bbe^\sharp= 2 \Delta^{-{1 \over 2}} {\hat{\bbe}}^\sharp, \qquad \bbe^\alpha = {1 \over \sqrt{2}} \Delta^{1 \over 4}
{\hat{\bbe}}^\alpha~.
\eea
In particular the metric on ${\cal S}$ written in terms of the new frame is
\bea
ds^2({\cal S})= 4 \Delta^{-1}\, (\hat\bbe^\sharp)^2+\Delta^{{1\over2}}\, \delta_{\a\bar\b}\, \hat\bbe^\a \hat\bbe^{\bar\b}~.
\eea
The geometric conditions (\ref{wg1})-(\ref{wg3}) on  ${\cal{S}}$ can now be rewritten as
\bea
d {\hat{\omega}}=0~,
\eea
\bea
-2 {\hat{\Omega}}_{\alpha, \beta}{}^\beta + {\hat{\Omega}}_{\sharp, \sharp \alpha} =0~,
\eea
and
\bea
 -{i \over 2} \bigg({\hat{\Omega}}_{\sharp, \alpha}{}^\alpha +{1 \over 2}{\hat{\Omega}}^{\alpha}{}_{, \sharp \alpha}
-{1 \over 2}{\hat{\Omega}}_{\alpha, \sharp}{}^\alpha \bigg)=1~,
\eea
where ${\hat{\omega}} = -i \delta_{\alpha \bar{\beta}} {\hat{\bbe}}^\alpha {\hat{\bbe}}^{\bar{\beta}}$, and $\hat\Omega$ is the spin connection computed in the
$\hat\bbe$ frame.

\subsubsection{Electric static $h=d\log \Delta$ horizons}

As for the $h=0$ solutions, if one furthermore imposes  $X=0$, then
additional conditions on the geometry are obtained. In particular, on taking the vector field dual to
${\hat{\bbe}}^\sharp$ to be ${\partial \over \partial \tau}$, one finds that ${\partial \over \partial \tau}$
is an isometry of ${\cal{S}}$, and $\Delta$ does not depend on $\tau$. Furthermore, on making an appropriate
$U(4)$ transformation on the holomorphic basis elements ${\hat{\bbe}}^\alpha$, one can without loss of generality
work with a $\tau$-independent basis
\bea
{\mathring \bbe}^\alpha = e^{- {i \tau \over 2}} {\hat{\bbe}}^\alpha\, \qquad {\mathring \bbe}^\sharp = {\hat{\bbe}}^\sharp~.
\eea

After some analysis of the KSEs, one finds
that ${\cal{S}}$ is a $U(1)$ fibration over an 8-dimensional compact K\"ahler  base manifold $B$ and the metric can be written as
\be
ds^2({{\cal{S}}}) = 4 \Delta^{-1} (d \tau  + \lambda)^2 + {1 \over 2} \Delta^{-{1 \over 2}} ds^2(B)~,
\ee
where  the $\tau$-independent
K\"ahler form{\footnote{Integrability of the almost complex
structure depends on the additional conditions obtained from setting $X=0$.}} is ${\hat{\omega}}$. The Ricci scalar and Ricci
form of $B$ satisfy
\bea
{\mathring R} =  \Delta^{-{3 \over 2}}, \qquad {\mathring \rho} =  2 d \lambda~,
\eea
and
\bea
{\mathring \nabla}^2 {\mathring R}  =   {1 \over 2} {\mathring R}^2 - {\mathring R}_{ij} {\mathring R}^{ij}~.
\la{rrr}
\eea
The 2-form Y is
\bea
Y = -3 \Delta^{-{3 \over 2}} (d \tau + \lambda) \wedge d \Delta - \Delta^{-{1 \over 2}} \bigg( {\mathring \rho}+{1 \over 2}
{\mathring R} {\mathring \omega} \bigg)
\eea
 So to construct such near horizon geometries, one has to find a 8-dimensional K\"ahler manifold such that the Ricci scalar
 and Ricci tensor satisfy (\ref{rrr}) with $\mathring R>0$. This equation has been obtained  \cite{kim} before in the search for gravitational duals
 in  AdS${}_2$/CFT${}_1$
 correspondence.

\newsection{Conclusions}

We have demonstrated that all static M-horizons are (local) warped products $\bR^{1,1} \times_w {\cal S}$ or
$AdS_2\times_w {\cal S}$, where ${\cal S}$ is a 9-dimensional manifold which admits either a $Spin(7)$ or $SU(4)$ structure
respectively, and the conditions on these structures imposed by supersymmetry have
been determined. If the M-horizons are electric and ${\cal S}$ has a
$Spin(7)$ structure, the near horizon geometry is $\bR^{1,1} \times {\cal S}$, ${\cal S}$ is a $Spin(7)$
holonomy manifold, and the 4-form flux vanishes.
However, for electric M-horizons such that ${\cal{S}}$ admits a $SU(4)$ structure, we have shown that
${\cal S}$ is a fibration
over a 8-dimensional K\"ahler manifold  $B$ whose Ricci scalar and Ricci tensor must satisfy
({\ref{rrr}}). This condition has also been found in \cite{kim} in the context of $AdS_2$
solutions in 11-dimensional supergravity.

It is remarkable that the classification of supersymmetric black hole horizons is closely related to that of Riemannian manifolds with special geometry. In the heterotic case, the understanding of horizons leads to a Calabi type of differential system on
conformally balanced  Calabi-Yau manifolds with torsion \cite{hh}. In the IIB case,
the horizons have sections which admit 2-strong Calabi-Yau structure with torsion \cite{iibh}. Furthermore as we have seen in M-theory, the existence of electric static horizons with $SU(4)$ structure  leads to a condition on the curvature of 8-dimensional
K\"ahler manifolds ({\ref{rrr}}). We remark that a special case of this condition arises when we take the
Ricci scalar of the 8-manifold to be constant.
Then the 8-manifold is K\"ahler-Yamabe, with the additional requirement that
the point-wise length of the Ricci tensor is constant.
More generally, it remains to determine the types of 8-dimensional compact K\"ahler manifolds, with positive
Ricci scalar, satisfying ({\ref{rrr}}). It is well-known that a K\"ahler metric $g$ can be deformed within its K\"ahler class and the deformation is determined by a single real function $f$ as $g\rightarrow g+i\partial\bar\partial f$. So starting from an arbitrary K\"ahler metric $g$, it may be possible to deform it such that the deformed metric satisfies 
 \bea
\label{mrrr}
(\nabla^2 R)_{g+i\partial\bar\partial f}=\Big({1 \over 2} { R}^2 - q(n) { R}_{ij} { R}^{ij}\Big)_{g+i\partial\bar\partial f}
\eea
for a unknown function $f$, where we have allowed K\"ahler manifolds of any dimension and so we have modified (\ref{rrr}) with a constant $q(n)$
which depends on the dimension $n$ of the K\"ahler manifold. 
 The subscript indicates that the Ricci scalar $R$, Ricci tensor $R_{ij}$, covariant derivative $\nabla$ and all the inner products are taken with respect to the deformed metric $g+i\partial\bar\partial f$.
There are many solutions to this equation but it is not apparent  that the general problem has always a solution.
 We remark that the RHS of ({\ref{mrrr}}) is non-negative if $q \leq 0$ and non-positive if
 $q \geq {n \over 2}$; in both cases compactness implies that the Ricci scalar is constant.
 In the case $q<0$, compactness implies the K\"ahler manifold is
 Ricci flat, and if $q=0$ the Ricci scalar vanishes. 
 If $q > {n \over 2}$ then compactness also implies that
 the manifold is Ricci flat. K\"ahler-Einstein manifolds satisfy ({\ref{mrrr}}) for $q={n \over 2}$,
 and Riemann surfaces satisfy ({\ref{mrrr}}) with $q=1$. For the case of interest for horizons with
 $n=8$ and $q=1$, the RHS of ({\ref{rrr}}) is of indeterminate sign.

Thus, for the systematic understanding of all horizons,  natural non-linear differential systems have to be solved on  compact manifolds with a special structure. The systematic investigation of solutions to such differential systems is an interesting problem in geometry which will have widespread applications in physics.

\vskip 0.5cm
\noindent{\bf Acknowledgements} \vskip 0.1cm
\noindent We thank James Lucietti for his constructive comments on static horizons.  GP thanks the PH-TH Divison at CERN for hospitality
where parts of this work were done. JG is supported by the EPSRC grant, EP/F069774/1.
GP is partially supported by the EPSRC grant EP/F069774/1 and the STFC rolling grant ST/G000/395/1.

\end{document}